\documentclass[a4paper]{jpconf}
\usepackage{graphicx}
\begin{document}
\title{On separate chemical freeze-outs of hadrons and light (anti)nuclei in high energy nuclear collisions}


\author{K. A. Bugaev$^{1}$, B. E. Grinyuk$^{1}$,  A. I. Ivanytskyi$^{1, 2}$,
         V. V. Sagun$^{1,3}$,  D. O. Savchenko$^{1}$, G. M. Zinovjev$^{1}$, 
        E. G. Nikonov$^{4}$,
              L. V.  Bravina$^{5}$,
        E. E.  Zabrodin$^{5, 6, 7}$,
         D. B.  Blaschke$^{7, 8, 9}$,
         S. Kabana$^{10, 11}$
        A. V.  Taranenko$^{7}$
}

\address{$^1$ Bogolyubov Institute for Theoretical Physics of the National Academy of Sciences of Ukraine, 03680 Kiev,  Ukraine}
\address{$^2$ Department of Fundamental Physics, University of Salamanca, Plaza de la Merced s/n 37008, Spain}
\address{$^3$ Centro de Astrof\'{\i}sica e Gravita\c c\~ao  - CENTRA,
Departamento de F\'{\i}sica, Instituto Superior T\'ecnico,
Universidade de Lisboa, 1049-001 Lisboa, Portugal}
\address{$^4$ Laboratory for Information Technologies, Joint Institute for Nuclear Research,   Dubna 141980, Russia}
\address{$^5$ Department of Physics, University of Oslo, PB 1048 Blindern, N-0316 Oslo, Norway}
\address{$^6$ Skobeltsyn Institute of Nuclear Physics, Moscow State University, 119899 Moscow, Russia}
\address{$^7$ National Research Nuclear University ``MEPhI'' (Moscow Engineering Physics Institute), 115409 Moscow, Russia}
\address{$^8$ Institute of Theoretical Physics, University of Wroclaw, pl. M. Borna 9, 50-204 Wroclaw, Poland}
\address{$^9$ Bogoliubov Laboratory of Theoretical Physics, JINR Dubna, Joliot-Curie str. 6, 141980 Dubna, Russia}
\address{$^{10}$ University of Nantes,  Facult\'e des Sciences et des Techniques,
2, rue de la Houssini\'ere, 44322 Nantes, France}
\address{$^{11}$ SUBATECH, Ecole des Mines, 4 rue Alfred Kastler, 44307 Nantes, France}
          
\ead{Bugaev@th.physik.uni-frankfurt.de}

\begin{abstract}
The multiplicities of light (anti)nuclei were measured recently by the ALICE collaboration in Pb+Pb collisions at the center-of-mass collision energy $\sqrt{s_{NN}} =2.76$ TeV. Surprisingly, the hadron resonance gas model  is able to perfectly describe  their multiplicities  under various assumptions. For instance, one can consider the (anti)nuclei  with a vanishing hard-core radius (as the point-like particles) or with the hard-core radius of proton, but the fit quality is the same for these assumptions. 
In this paper we assume the hard-core radius of nuclei consisting of $A$ baryons or antibaryons to follow the simple law 
$R(A) = R_b (A)^\frac{1}{3}$, where  $R_b$ is the hard-core radius of nucleon. 
To implement such a relation into the hadron resonance gas model we employ the induced surface tension concept  and  analyze  the  hadronic and (anti)nuclei multiplicities measured by the ALICE collaboration. The hadron resonance gas model with the induced surface tension allows us to verify different scenarios of chemical freeze-out of (anti)nuclei. It is shown that the most successful description of hadrons can be achieved at the chemical freeze-out temperature $T_h=150$ MeV, while the one for all (anti)nuclei is $T_A=168.5$ MeV.  Possible explanations of this  high temperature of (anti)nuclei chemical freeze-out are discussed.
\end{abstract}

\section{Introduction}
\label{KAB_sectI}

Light (anti)nuclei production measured recently by the ALICE collaboration in Pb+Pb collisions at the center-of-mass collision energy $\sqrt{s_{NN}} =2.76$ TeV \cite{KAB_Ref1}  caught a considerable attention. Partly it is inspired by the fact that it looks rather surprising that light nuclei with binding energies of an order of a few MeV are produced at all in such violent collisions.  
Two main approaches, thermal production  model \cite{KAB_Ref2,KAB_Ref3,KAB_Ref3b,KAB_Ref4,KAB_Ref5} and coalescence one \cite{KAB_Ref6,KAB_Ref7,KAB_Ref8,KAB_Ref9,KAB_Ref11,KAB_Ref12} are able to  explain  this phenomenon equally well. Usually,  the thermal model assumes a perfect chemical equilibrium above the chemical freeze-out (CFO) temperature   $T_{CFO}$ and a sharp CFO of all hadrons at this temperature. After the CFO  the yields are assumed to be unchanged, but particles can scatter elastically until the system reaches the moment of kinetic freeze-out,  which
at the ALICE experiment was reported to occur at temperature about 115 MeV \cite{KAB_Ref1}, while the corresponding  CFO temperature varies  found in thermal model varies from $150$ \cite{KAB_Ref4,KAB_Ref5}
MeV to 160 MeV \cite{KAB_Ref2}. 
In contrast to these assumptions, the coalescence approach postulates that the light nuclei are formed only at  late times of the nuclear-nuclear reaction via the  binding of  nucleons that move close in phase space. 

The main problem with the existing formulations of  thermal model is that none of them is suited to treat the excluded volume of (anti)nuclei 
and hadrons on the same footing. Indeed,  in  light (anti)nuclei  the nucleons and hyperons are loosely bound  and, hence, 
the radius of  small nuclei of $A$ nucleons, or baryons in general,  with $2\le A \le 4$  is about 
1.8-2 fm \cite{KAB_Ref13}, 
 i.e. it is essentially  larger than the maximal double hard-core radius of hadrons ($0.84$ fm) \cite{KAB_Ref4,KAB_Ref5} even for  a deuteron ($A=2$).   The total proper volume of  $A$ baryons with the hard-core radius $R_b$  is  $V_A = A \frac{4}{3}\pi R_b3$. Hence, the equivalent  hard-core radius of  a nucleus consisting of $A$ baryons   is
 \begin{equation}\label{KAB_EqI}
 R_{A} =  A^\frac{1}{3}R_b\,, 
\end{equation}
since each baryon in a nucleus  interacts with external particle individually and the other baryons do not affect them due to a very large distance
between baryons inside a nucleus. 
Note that the recently developed hadron resonance gas model (HRGM) which is based on the induced surface tension (IST) concept \cite{KAB_Ref4,KAB_Ref5,KAB_Ref14} belongs to the class of HRGM (see, e.g., \cite{Albright:2014gva,Typel:2016srf}) 
which can handle any number of individual hadron excluded volumina corresponding to different hard-core radii 
(for  a discussion and important applications see also   \cite{Typel:2017vif}). 
Hence, in this work we report our preliminary results on the analysis of the ALICE hadronic and (anti)nuclei multiplicities measured at the collision energy $\sqrt{s_{NN}} =2.76$ TeV with the hard-core radii of nuclei given  by Eq.  (\ref{KAB_EqI}).  Since the list of 
analyzed data is rather long, but is  well-known,  we refer to the original works \cite{KAB_Ref3b,KAB_Ref4,KAB_Ref5} in which these  ALICE data were thoroughly analyzed.  Our  treatment of hadronic data follows the line of Refs.  \cite{KAB_Ref4,KAB_Ref5},
i.e. the ratios will be analyzed, while the (anti)nuclei  multiplicities are included into a fitting procedure in a way which  allows us to verify two different hypotheses  about their production. 

The work is organized as follows.  In Sect. \ref{KAB_sect2} we describe the main features of the HRGM with IST. The results of the ALICE data analysis are presented in Sect.  \ref{KAB_sect3}, while our conclusions are formulated in Sect.  \ref{KAB_sect4}.

\section{HRGM with multicomponent hard-core repulsion}
\label{KAB_sect2}

The most advanced version of the HRGM with IST \cite{KAB_Ref4,KAB_Ref5} was originally formulated for the Boltzmann particles  \cite{KAB_Ref14}, but recently it is straightforwardly generalized to the cases of quantum particles \cite{KAB_Ref15,KAB_Ref16}. 
For the   high CFO  temperatures achieved in Pb+Pb reactions one can safely use the Boltzmann statistics for all hadrons and (anti)nuclei \cite{KAB_Ref4,KAB_Ref5}. Moreover, this provides an essential speed up of the fitting process, since the momentum integration can be done only  once for each particle species. 

In the grand canonical ensemble the HRGM with IST   can be written as  a system of two coupled  equations for the system pressure $p$ and the 
IST coefficient  $\Sigma$: 
\begin{eqnarray}
\label{EqII}
p = 
  T \sum_{k=1}^N \phi_k \exp \left[ 
\frac{ - v_k p - s_k \Sigma}{T} \right]
\,, \quad 
\Sigma =
 T \sum_{k=1}^N R_k \phi_k \exp 
\left[ \frac{- v_k p - s_k \alpha \Sigma}{T} \right] \,,
\end{eqnarray}
where  $v_k = \frac{4}{3}\pi R_k^3$ is the proper volume of the particle with hard-core radius $R_k$,  $s_k = 4\pi R_k^2 $
denotes its proper surface and 
 $\alpha = 1.25$ \cite{KAB_Ref4,KAB_Ref5}. 
 For the (anti)nuclei of $A$ (anti)baryons  the hard-core radius is given by Eq. (1). 
  In  the system  (\ref{EqII})  all chemical potentials are set to zero, since 
 at this value of collision energy there  is almost no difference between particle and antiparticle \cite{KAB_Ref3b,KAB_Ref4,KAB_Ref5}.
The thermal density of  the $k$-th particle  species accounts for the Breit-Wigner mass attenuation of hadron resonances 
\vspace*{-1.1mm}
\begin{eqnarray}
\label{EqIV}
\phi_k = g_k  \hspace*{-5.5mm} && \int\limits_{M_k^{Th}}^\infty  \,  \frac{ d m}{N_k (M_k^{Th})} 
\frac{\Gamma_{k}}{(m-m_{k})^{2}+\Gamma^{2}_{k}/4}
\int \frac{d^3 p}{ (2 \pi)^3 }   \exp \left[ -\frac{ \sqrt{p^2 + m^2} }{T} \right] \,,
\end{eqnarray}
where $m_k$ and  $\Gamma_k$  denote, respectively,  the mass and  width   of the $k$-th particle  species. 
The factor 
${N_k (M_k^{Th})} \equiv \int\limits_{M_k^{Th}}^\infty \frac{d m \, \Gamma_{k}}{(m-m_{k})^{2}+\Gamma^{2}_{k}/4} $ denotes 
a corresponding normalization, while $M_k^{Th}$ corresponds to the decay threshold mass of the $k$-sort of hadrons in the leading channel.
Although  Eq. (\ref{EqIV}) is an approximation for the wide resonances, it is known that such an approximation is sufficiently 
accurate  \cite{KABlec_Kuksa}. It is necessary  to stress that the  mass attenuation  in Eq.  (\ref{EqIV})
for a mixture of  hadron resonances can be rigorously obtained   \cite{KAB_David:16A,KAB_David:16B} from  the Phi-functional approach \cite{KAB_Phi-approach},  when the
Phi-functional is chosen from the class of  two-loop diagrams only, see also \cite{KAB_Phi-approach3}.

In contrast to the usual multicomponent HRGM formulations to 
determine the particle number densities $\{ \rho_k \}$ one needs to 
solve only a system of  two equations of the system  (2)  (the total strange charge is zero by construction)
irrespective to the number of different hard-core radii in the model.
Hence, we believe that HRGM with IST  is 
perfectly suited for the analysis of all hadronic multiplicities 
which will be measured in the nearest future at SPS, RHIC, NICA and FAIR. 

Another great advantage of the HRGM with IST   is its validity up to the 
packing fractions $\eta \equiv \sum_k \frac{4}{3}\pi R_k^3 \rho_k 
\simeq  0.2-0.22$ \cite{KAB_Ref4,KAB_Ref5}, i.e.  at the particle number 
densities for which the traditional HRGM  based on the Van der Waals 
approximation is entirely wrong.  
Using the particle number density $\rho_k$ of $k$-th sort of  hadrons 
 one can determine the thermal $N_k^{th} =V \rho_k$ ($V$ is the 
effective volume at CFO) and the total multiplicities $N_k^{tot}$.
The latter should account for the hadronic decays after the CFO and, hence, 
 the ratio of total hadronic multiplicities becomes 
\begin{equation}
\label{EqIX}
\frac{N^{tot}_k}{N^{tot}_j}=
\frac{\rho_k+\sum_{l\neq k}\rho_l\, Br_{l\rightarrow k}}{\rho_j+
\sum_{l\neq j}\rho_l \, Br_{l\rightarrow j}}\,,
\end{equation}
where $Br_{l\rightarrow k}$ is the branching ratio, i.e. a probability 
of particle $l$ to decay strongly into a particle $k$. More details on 
the fitting procedure of experimental data with the HRGM can be found in  
\cite{KAB_Ref4,KAB_Ref5}.

\section{Results}
\label{KAB_sect3}

The HRGM with IST outlined above is applied to describe the whole set of the ALICE data measured at $\sqrt{s_{NN}} =2.76$ TeV.
First we consider the traditional approach to CFO in which all particles, i.e. hadrons and (anti)nuclei, have a single CFO hyper-surface (Model I hereafter).
The total $\chi^2_{tot}(V)$ of Models I and II is defined as 
\begin{eqnarray}\label{Eq_VI}
\chi^2_{tot}(V) &=& \chi^2_{h} + \chi^2_{A} (V)
= \sum_{k\in h} \left[  \frac{R_k^{theo} - R_k^{exp}}{\delta R_k^{exp}}\right]^2  +   \sum_A \left[  \frac{\rho_A(T) V - N^{exp}_A}{\delta N^{exp}_A}\right]^2  \,,
\end{eqnarray}
where $\chi^2_{h}$ and $\chi^2_{A}$ denote the mean deviation squared for hadrons and (anti)nuclei, respectively. 
Since hadronic part of $\chi^2_{tot}(V)$ involves only the ratios of hadron multiplicities  \cite{KAB_Ref4,KAB_Ref5}, then it does not contain
the CFO volume $V$. On a contrary, the  (anti)nuclei part of $\chi^2_{tot}(V)$ depends on the thermal densities of (anti)nuclei of $A$ (anti)baryons and the CFO volume.  The main assumption of the Model I is that  the CFO volume is defined by the multiplicity of $\pi^+$-mesons as
$V_I = \frac{N^{exp}_{\pi^+}}{\rho_{\pi^+}}$.  Hence, the Model I has two  parameters  to reproduce data, namely CFO temperature and $V_I$.  We tried another way of fitting by introducing the ratios for (anti)nuclei to the multiplicity of 
$\pi^+$-mesons, like it is done for hadronic ratios. However, in these case one automatically increases the resulting error for the ratio, although we got practically the same results of fit like for the Model I. The only found difference between this  prescription from the Model I is that the $\chi^2_{tot}(V)$  minimum is slightly  broader and slightly lower, but the CFO temperature is  the same as in the Model I. 

The Model I provides a typical poor quality  of the ALICE data description with  $\min \{ \chi^2_{tot} (V_I) \} \simeq 30.77 $,  $\chi^2_{1h}  \simeq 17.69$ and  $\chi^2_{1A} (V_I)  \simeq 13.08$ which results 
in $\min \{ \chi^2_{tot} (V_I)/dof \} \simeq 30.77/17 \simeq 1.81$, if one uses 
the hard-core radius of (anti)nuclei according to Eq. (1). The CFO temperature $T_{CFO} \simeq 157.1 \pm 6$ MeV and volume $V_I \simeq 7467$ fm$^3$ and the fit quality of the Model I are very similar to the results of Ref.  \cite{KAB_Ref3b}.   A poor quality of the Model I description is rooted in the fact that the local minimum of the hadronic $\chi^2_{1h}$  is located at the temperature about $150$ MeV, while the local minimum of  $\chi^2_{A} (V_I)$ for   (anti)nuclei  is located at the temperature about $161.5$ MeV. Therefore the minimum of  their sum is located in between them and provides  essentially worse quality than either of its parts. 

The  Model II differs from the Model I by an assumption that CFO of (anti)nuclei occurs at the other hyper-surface.  Note that a few years ago
two groups independently suggested an idea that some hadronic flavors can experience the CFO separately from the other ones due to a different 
underlying mechanism of their freeze-out \cite{KABSFO, KAB_Ref23, KAB_Ref24,KAB_Ref25,KAB_Ref26}. The Model II generalizes such a hypothesis to the (anti)nuclei and, hence, the CFO volume $V_{II}$ in this case  is found from the minimum of  $\chi^2_{tot} (V)$ with respect to  $V$:
\begin{eqnarray}\label{Eq_VII}
\frac{d \chi^2_{tot} (V)}{d V} = 0 \quad \Rightarrow \quad V_{II} (T) = {\sum_A  \frac{\rho_A(T) N^{exp}_A}{[\delta N^{exp}_A]^2}}\cdot \left[ \sum_A  \frac{\rho_A(T) \rho_A(T)}{[\delta N^{exp}_A]^2} \right]^{-1} \,.
\end{eqnarray}
Substituting $V_{II} (T)$ of Eq.  (\ref{Eq_VII}) into expression for  $\chi^2_{A} (V)$ from Eq. (\ref{Eq_VI}),
one can find the global minimum of $\chi^2_{A} (V_{II} (T))$  with respect to CFO temperature $T$.  In this case the CFO temperature of (anti)nuclei is $T_A=168.5$ MeV, i.e. it is essentially higher than the one for hadrons $T_h =150$ MeV, whereas the CFO volume 
$V_{II} = 2725$ fm$^3$ is smaller than the corresponding volume of hadronic CFO $V_h = 8965$  fm$^3$.  
Another striking result of  Model II  is that the separate description of hadronic and (anti)nuclei data provides a sizably better  quality than the one obtained by the  Model I and other versions of HRGM, i.e. $\chi^2_{tot} (V_{II} (T)) \simeq 13.27/16 \simeq 0.83$ with a single additional
fitting parameter  compared to the Model I. 

From the values of CFO volumes obtained by  Model II one may deduce that the (anti)nuclei are produced from the  quark gluon plasma bags earlier than hadrons. Such a conclusion may partly explain the fact that the (anti)nuclei can survive during the expansion of these bags and their subsequent hadronization.
If the initial collective velocity of (anti)nuclei is sufficiently high, than the hadrons will be able to interact and destroy them at very late times of 
heavy ion collision reaction. 

However, our educated guess is that the high CFO temperature of (anti)nuclei is caused by the fact that the quark gluon bags formed in Pb+Pb collisions  have  a mass spectrum  of the Hagedorn model \cite{KAB_Ref27}. In this case the temperature of the hadrons produced from such bags depends not only 
on the masses of emitted particles, their number and the mass of  quark gluon bag \cite{KAB_Ref28,KAB_Ref29}, but also on their mass spectrum \cite{KAB_Ref29}.  If  the mass spectrum of quark gluon bags is not purely exponential, but has a pre-factor $m^{-3}$ 
like in the  Hagedorn model, than using the formalism of Ref. \cite{KAB_Ref29} one can show that temperature of many light particles (hadrons) produced from such  bags will be  10-15\% smaller than the one of a few  heavy particles (nuclei). Note that these difference  of  temperatures is very similar to our findings for the Model II. 
Furthermore, in \cite{SKPMosci} it is discussed also the possibility that the hadron mass spectrum 
(including the experimentally missing states) may not be exponential.

\begin{figure}[t]
\centering
\mbox{\includegraphics[width=80mm,clip]{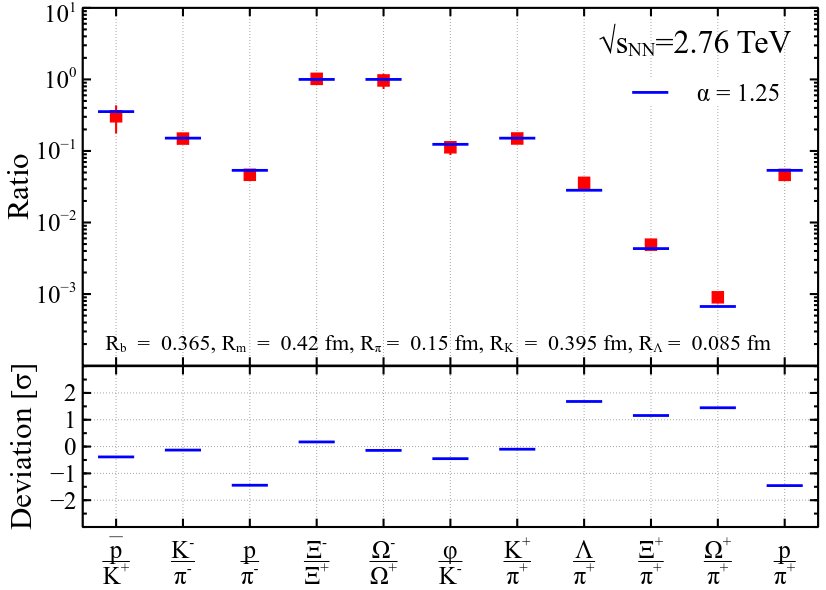}
\hspace*{-2.2mm}
\includegraphics[width=80mm,clip]{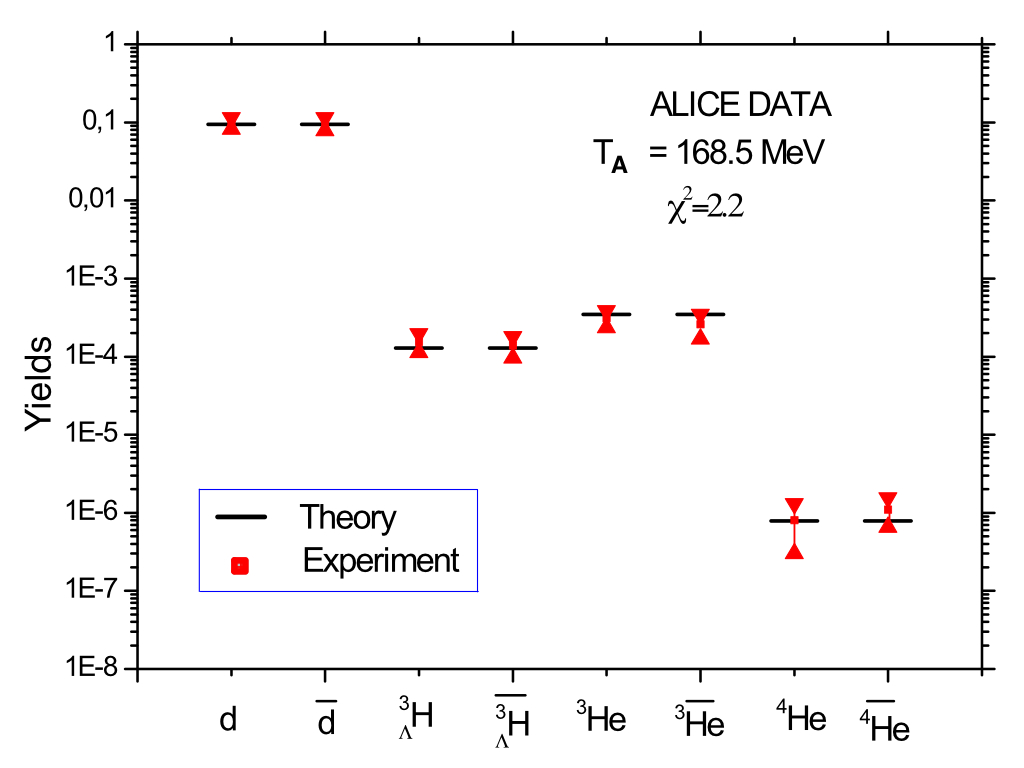}}
\caption{{\bf Left panel:} The hadronic ratios  from Ref. \cite{KAB_Ref4} were fitted by the HRGM with IST for the set of hard-core radii shown in the left panel. 
 The obtained  CFO temperature for hadrons is  $T_{CFO} \simeq 150 \pm 7$ MeV. 
 The  quality  of the fit is  $\chi^2_{2h}/dof \simeq 11.07/10 \simeq 1.1$. The upper panel shows the fit of the ratios, while the lower panel shows the deviation between data and theory in the units of  error.
{\bf Right panel:}  The (anti)nuclei yields found by the ALICE experiment (symbols) are compared with the ones obtained by the HRGM with IST for the scenario of separate CFO of nuclei (bars). The fit quality (anti)nuclei yields is $\chi^2_{2A}/dof \simeq 2.2/6 \simeq 0.367 $. 
}
\label{KAB_fig2}       
\end{figure}  

\section{Conclusions}
\label{KAB_sect4}

In this work we developed  the hadron resonance model with the induced surface tension which allows us to treat the hard-core repulsion of hadrons and light (anti)nuclei on the same footing.  In contrast to all previous studies here we employ the hard-core radius of $A$ nucleons in the form 
$R(A) = R_b (A)^\frac{1}{3}$, where $R_b = 0.365$ fm denotes the hard-core radius of (anti)baryons.  Then we consider two models of chemical freeze-out of (anti)nuclei. The first of them, Model I, corresponds to the usual assumption that the hadrons and (anti)nuclei multiplicities are frozen at  the same hyper-surface. In this case we obtain the typical value of the fit quality $\min \{ \chi^2_{tot} (V_I)/dof \} \simeq 30.77/17 \simeq 1.81$ which is slightly  better than the results of Ref. \cite{KAB_Ref3b}.  The Model II assumes that the chemical freeze-out 
of nuclei occurs separately from the one of hadrons.  In this case we have one extra fitting parameter compared to the Model I, namely the temperature of (anti)nuclei $T_A = 168.5$ MeV, which is essentially higher than the one for hadrons $T_h = 150$ MeV, but is 
similar to the one found for the highest RHIC energies \cite{KAB_Ref4,KAB_Ref5}.  
Such a model provides  $\min \{ \chi^2_{tot} (V_II)/dof \} 13.27/16 \simeq 0.83$ which is more than two times smaller than the corresponding value of the Model I.
 We believe that a possible explanation for such a high value of the chemical freeze-out temperature of (anti)nuclei may be related  to the fact that  
 the quark gluon bags producing the (anti)nuclei   have not purely exponential mass spectrum $\exp{(m/T_H)}$, but the one 
  which  corresponds to the Hagedorn model, i.e. $m^{-3}\, \exp{(m/T_H)}$.  In this case the temperature of  light and heavy particles produced by such bags may differ  by 10-15\%, which is close to our results for the Model II.

\ack{\small
The work of K.A.B., 
A.I.I., V.V.S., B.E.G., D.O.S. and G.M.Z. was supported 
by the Program of Fundamental
Research of the Department of Physics and Astronomy of the National
Academy of Sciences of Ukraine (project No. 0117U000240).
V.V.S. thanks the 
Funda\c c\~ao para a Ci\^encia e Tecnologia (FCT), Portugal, for the
financial support through the grant No.
UID/FIS/00099/2013 to make research at the CENTRA, Instituto Superior T\'ecnico, Universidade de 
Lisboa. 
The work of L.V.B. and E.E.Z. was supported by the Norwegian 
Research Council (NFR) under grant No. 255253/F50 - CERN Heavy Ion 
Theory. 
L.V.B. and K.A.B. thank the Norwegian Agency for International Cooperation and Quality Enhancement in Higher Education for financial support, grant 150400-212051-120000 "CPEA-LT-2016/10094 From Strong Interacting Matter to Dark Matter". 
D.B.B. acknowledges a support from the Polish National Science Centre (NCN) under grant no. UMO-2014/15/B/ST2/03752.
The work of A.V.T. 
was partially supported by the Ministry of Science and Higher Education
of the Russian Federation, grant No 3.3380.2017/4.6.
D.B.B. and A.V.T. acknowledge partial support from the National Research Nuclear University ``MEPhI'' in the framework of the Russian Academic Excellence Project (contract no. 02.a03.21.0005, 27.08.2013).  
The work of A.I.I. was done within the project SA083P17 of Universidad de Salamanca launched by the Regional Government of Castilla y Leon and the European Regional Development Fund. 
The authors are grateful to the COST Action CA15213 ``THOR" for supporting their networking activities. 
}

\end{document}